\documentclass[12pt]{article}
\usepackage{amsfonts}
\usepackage{amssymb}
\usepackage{amsbsy}
\usepackage{graphics}
\usepackage{amsmath}
\usepackage{multirow}
\usepackage{pgf,pgfarrows}

\input epsf.sty

\newlength{\TZ}
\newcommand{\BEQ}{\begin{equation}}     
\newcommand{\BEA}{\begin{eqnarray}}
\newcommand{\BD}{\begin{displaymath}}
\newcommand{\EEQ}{\end{equation}}       
\newcommand{\EEA}{\end{eqnarray}}
\newcommand{\ED}{\end{displaymath}}
\newcommand{\D}{{\rm d}}                
\newcommand{\II}{{\rm i}}               
\newcommand{\wit}[1]{\widetilde{#1}}    

\renewcommand{\vec}[1]{\boldsymbol{#1}} 

\newcommand{\appsection}[2]{\setcounter{equation}{0} \section*{Appendix #1. #2}
\renewcommand{\theequation}{#1\arabic{equation}}
              \renewcommand{\thesection}{#1} }

\catcode`\@=11
\def\numberbysection{\@addtoreset{equation}{section}
        \def\theequation{\thesection.\arabic{equation}}}
\numberbysection

\begin{document}

\begin{titlepage}

\vskip 1.5 cm
\begin{center}
{\Large \bf {\tt } 
Ageing in bosonic particle-reaction models with long-range transport}
\end{center}

\vskip 2.0 cm
\centerline{ {\bf Xavier Durang} and {\bf Malte Henkel} }
\vskip 0.5 cm
\centerline {D\'epartement de Physique de la Mati\`ere et des Mat\'eriaux,}
\centerline{Institut Jean Lamour\footnote{Laboratoire associ\'e au CNRS UMR 7198}, CNRS -- Nancy Universit\'e -- UPVM,}
\centerline{B.P. 70239, F -- 54506 Vand{\oe}uvre l\`es Nancy Cedex, France}

\begin{abstract}
Ageing in systems without detailed balance is studied in bosonic contact and 
pair-contact processes with L\'evy diffusion. In the ageing regime, the 
dynamical scaling of the two-time correlation function and two-time response 
function is found and analysed. Exact results for non-equilibrium exponents and scaling 
functions are derived. The behaviour of the fluctuation-dissipation ratio is 
analysed. A passage time from the quasi-stationary regime to the ageing regime is defined, in qualitative agreement with kinetic spherical models and $p$-spin spherical glasses.
\end{abstract}

\vfill
\end{titlepage}
\setcounter{page}{2}

\section{Introduction}

One of the paradigmatic example of non-equilibrium critical behaviour is furnished by ageing systems. A common way to realise physical ageing is to move a system rapidly out of equilibrium by `quenching' some thermodynamic parameter into a coexistence region of its phase diagramme such that there exist several competing thermodynamical states. Alternatively, one may quench a system exactly onto a critical point of its stationary state, which is the case we shall study in this paper. From a phenomenological point
of view, physical ageing may by characterised by the three properties of (i) slow, i.e.
non-exponential dynamics, (ii) breaking of time-translation-invariance and (iii)
dynamical scaling.

Although physical ageing was first recognised and studied in glassy systems, starting from
Struik's classical experiments \cite{Stru78,Vinc07}, the analysis of many aspects of ageing
is more simple in non-glassy systems without disorder or frustrations.\footnote{A careful comparison of the ageing properties of glassy and non-glassy systems and a detailed
appreciation of their differences was given by
Chamon, Cugliandolo and Yoshino \cite{Cham06}.} Typical examples of this kind are simple ferromagnets,
the dynamics of which can be characterised in terms of a single time-dependent length
scale $L(t)\sim t^{1/z}$, which defines the dynamic exponent, see \cite{Bray94,Cala04}
for reviews. Their ageing behaviour is conveniently studied through the two-time
correlation and response functions of the time-dependent order-parameter $\phi(t)$, for
which ones expects the scaling behaviour
\BEA
C(t,s) &=& \langle\phi(t)\phi(s)\rangle-\langle\phi(t)\rangle \langle \phi(s)\rangle \:= \: s^{-b} f_{C}(t/s)  
\label{CR1} \\
R(t,s)&=& \left.\frac{\delta\langle\phi(t)\rangle}{\delta h(s)}\right|_{h=0} \:= \: s^{-1-a} f_{R}(t/s) 
\label{CR2}
\EEA
where $h(s)$ is an external magnetic field conjugate to the order-parameter, and 
$t,s$ are the observation and waiting times, respectively. This scaling behaviour should
be valid in the ageing regime $t,s\gg \tau_{micro}$ and $t-s \gg \tau_{micro}$ where $\tau_{micro}$ is a microscopic reference time. From the asymptotic behaviour 
$f_{C,R}(y)\sim y^{-\lambda_{C,R}/z}$ of the scaling functions as $y\to\infty$ one may 
define the autocorrelation and autoresponse exponents $\lambda_C$ and $\lambda_R$, respectively. The exponents $a,b$ are known as ageing exponents and can be expressed in terms
of $z$ and static critical exponents. For reviews, see e.g. \cite{Bray94,Cugl02,Godr02,Cala04,Henk07b}. 

In this paper, we are interested in systems where the underlying dynamics does not
satisfy detailed balance such that the stationary state is no longer an equilibrium state. 
Recently, it was shown that the same kind of non-equilibrium scaling behaviour 
(\ref{CR1},\ref{CR2}) 
as described above for ageing ferromagnets applies to this kind of systems, as exemplified
for the critical contact process \cite{Enss04,Rama04,Baum07} and the critical 
non-equilibrium kinetic Ising model \cite{Odor06}. However, in contrast to the non-equilibrium critical dynamics of ferromagnets, the ageing exponents $a$ and $b$ need no longer to be equal, which for the contact process can be understood as a consequence of its model-specific rapidity-reversal-symmetry \cite{Baum07}. Furthermore, there are exactly solvable models such as the so-called `bosonic' contact and pair-contact processes\footnote{The terminology `bosonic' refers to the property of these models 
(referred to as BCPD and BPCPD, respectively) that an arbitrary number of
particles per site is allowed, in contrast to the usual contact or pair-contact processes which obey the `fermionic' constraint of at most one particle per site.} \cite{Houc02,Paes04a,Paes04b} for which the ageing behaviour could be analysed in full detail and the non-equilibrium exponents and scaling functions be calculated exactly \cite{Baum05}. The critical behaviour of the
`bosonic' models is different from the more usually studied `fermionic' ones since, although
the average total number of particles is conserved at criticality, for
long times the particles may condense onto a single site which mimics the inhomogeneous
growth of bacteria colonies and had originally led to the formulation of the BCPD \cite{Youn01,Houc02} and BPCPD. All these models show
ageing only when brought to the critical point(s) of the stationary states, 
see \cite{Henk07} for a recent review. They also have in common that transport of single
particles is only through the `diffusive' hopping to nearest-neighbour sites. 
We wish to investigate the consequences when that local, nearest neighbour transport of single particles is replaced by long-range L\'evy flights, leading to superdiffusive behaviour. There are several motivations for such an undertaking:
\begin{enumerate}
\item L\'evy flights \cite{Shle95, Metz00} generalise the usually considered local (brownian) random walks in that they lead to a generalised central limit theorem for the sum of a 
large number of independent random variables. Heuristically, it appears reasonable to 
consider the `particles' which make up the basic dynamics of non-equilibrium systems as 
coarse-grained variables which arise from summing over many more microscopic degrees of freedom. It may hence appear natural that the effects of 
long-range jumps between distant sites should be taken into consideration. This might become relevant for the description of traces particles in turbulent flows or the spreading of epidemics, see \cite{Lube} and refs there in.
\item When studying the ordinary contact process with L\'evy flights, it has been shown
that the critical behaviour of the stationary state, as well as the relaxation behaviour, 
is modified \cite{Hinr99,Jans99,Delo02,Bara09,Henk08}. 
\end{enumerate}

The models under study are defined as follows. Each site of an infinite $d$-dimensional hyper-cubic lattice may contain an arbitrary positive number of particles. Particles on the same site can undergo the reactions 
\BEA \label{gl:reac}
mA &\longrightarrow & (m+k)A    \hspace{2.0 truecm}\mbox{\rm with rate $\mu$} 
\nonumber \\
mA &\longrightarrow & (m-\ell)A \hspace{2.0 truecm}\mbox{\rm with rate $\lambda$} 
\EEA
Furthermore, single particles can hop to another site at distance $|\vec{r}|$, with a
rate 
\begin{equation} \label{gl:diff}
D({\bold r})= \frac{D}{(2\pi)^{d}} \int_{-\infty}^{\infty} \!\D^{d}{\bold q}\: 
e^{\left({\rm i}{\bold q} \cdot {\bold r}-c\| {\bold q}\|^{\eta} \right)}
\end{equation}
where $0<\eta < 2$ is a control parameter and $c$ is a non-universal dimensionful constant. The case $m=1$ (and with $k=\ell=1$) is called the {\em bosonic contact process with L\'evy flight} (BCPL) and the case $m=2$ is called the {\em bosonic pair-contact process with L\'evy flight} (BPCPL).\footnote{The results previously found for the BCPD and BPCPD \cite{Houc02,Paes04a,Baum05} will be recovered in the limit case $\eta \rightarrow 2$.}

In section 2, we write down the closed set of equations of motion for the correlation and response functions. In section 3, we discuss the phase diagram, exponents and scaling functions for both models. In section 4, we verify that the exact scaling functions found for the bosonic contact process can be understood from the theory of local scale-invariance. A discussion of local scale-invariance for bosonic pair-contact process with L\'evy diffusion will be presented in a sequel paper. In section 5, we study the passage from the quasi-stationary regime to the ageing regime
and identify the passage exponent $\zeta$ which characterises the relevant time-scale $\tau_p(s)\sim s^{\zeta}$ \cite{Zipp00}, for the first time in a critical system.
In section~6, we present our conclusions. Finally, we derive in appendix~A a relation for the physical interpretation 
of the two-time correlators, list in appendix~B the scaling functions for the space-time correlators and responses
and present in appendix~C the analysis of the passage time in the critical spherical model. 

\section{Equations of motion}

Following \cite{Paes04a}, the master equation is written in 
the quantum hamiltonian/Liouvillian formulation (for reviews, see \cite{Schu00,Henk08}) 
as $\partial_{t}\vert P(t)\rangle=-H\vert P(t)\rangle$ where $P(t)$ is the 
time-dependent state vector and the hamiltonian can be expressed in terms of annihilation and creation operators $a({\bold x})$ and $a^{\dagger}({\bold x})$. We also define the particle number operator as $n({\bold x})=a^{\dagger}({\bold x})a({\bold x})$. The Hamiltonian reads
\begin{equation}
\begin{split}
H = &-\sum_{{\bold r}\neq {\bold 0}} \sum_{{\bold x}} D({\bold r})[a({\bold x})a^{\dagger}({\bold x}+{\bold r})-n({\bold x})]\\
&-\lambda\sum_{{\bold x}}[a^{\dagger}({\bold x})^{m-\ell}a({\bold x})^{m}-\prod_{i=1}^{m}(n({\bold x})-i+1)]\\
&-\mu\sum_{{\bold x}}[a^{\dagger}({\bold x})^{m+k}a({\bold x})^{m}-\prod_{i=1}^{m}(n({\bold x})-i+1)]-\sum_{{\bold x}}h({\bold x})a^{\dagger}({\bold x})
\end{split}
\end{equation}
For the computation of the response function, we also added an external field which describes the spontaneous creation of a single particle with a site-dependent rate $h=h({\bold x})$ on the site ${\bold x}$. 

Single-time observables $g(t,{\bold x})$ can be obtained from the time-independent quantities by switching to the Heisenberg picture. The differential equations for the desired quantities can be obtained by using the usual Heisenberg equation of motion $\partial_t g=\left[g,H\right]$. The space-time dependent particle density $\rho(t,{\bold x})=\langle a^{\dagger}(t,{\bold x}) a(t,{\bold x})\rangle=\langle a(t,{\bold x}) \rangle$ satisfies
\begin{equation}
\begin{split}
\frac{\partial}{\partial t}\langle a(t,{\bold x})\rangle= &\sum_{{\bold r}\neq {\bold 0}} D({\bold r})[\langle a(t,{\bold x}+{\bold r})\rangle-\langle a(t,{\bold x})\rangle]\\
&-\lambda \ell\langle a(t,{\bold x})^{m}\rangle +\mu k\langle a({t,\bold x})^{m}\rangle +h({t,\bold x})
\end{split}
\label{eqmvt1}
\end{equation}
For the bosonic contact process BCPL, this equation closes for arbitrary values of the rates. However, for the bosonic pair-contact process BPCPL, we only find a closed set of equations along the critical line defined as 
\begin{equation}
\sigma = \frac{\mu k - \lambda \ell}{ D} \stackrel{!}{=} 0
\label{criticalline}
\end{equation}
As we shall see later, along the critical line the space-integrated particle density
\BEQ
\rho_0 := \int \!\D\vec{x}\: \rho(t,\vec{x}) = \int \!\D\vec{x}\: \langle a(t,\vec{x})\rangle
\EEQ
is conserved, although the microscopic process can change the total number of particles in the
system. All this is completely analogous to what was already known for the BCPD and BPCPD. 

Throughout this paper, we are interested in the correlation and response functions
\BEA
C(t;\vec{r}) &=& \langle a(t,\vec{x}) a(t,\vec{x} + \vec{r}) \rangle - \rho_0^2
\nonumber \\
C(t,s;\vec{r}) &=& \langle a(t,\vec{x}) a(s,\vec{x} + \vec{r}) \rangle - \rho_0^2
\label{gl:CRdef} \\
R(t,s;\vec{r}) &=& \left.\frac{\delta\langle a(t,\vec{x}+\vec{r})\rangle}{\delta h(s,\vec{x})}
\right|_{h=0}
\nonumber
\EEA
In these notations, we already anticipate spatial translation-invariance. 
The single-time correlator is determined from the following equations of motion
\begin{equation}
\begin{split}
\frac{\partial}{\partial t}\langle a(t,{\bold x})a(t,{\bold x})\rangle= &\,2\sum_{{\bold r}\neq {\bold 0}} D({\bold r})[\langle a({t,\bold x})a({t,\bold x}+{\bold r})\rangle-\langle a(t,{\bold x})^{2}\rangle]\\
&+\lambda \ell [(1+\ell-2m)\langle a(t,{\bold x})^{m}\rangle-2\langle a(t,{\bold x})^{m+1}\rangle]\\
&-\mu k [(1-k-2m)\langle a(t,{\bold x})^{m}\rangle-2\langle a(t,{\bold x})^{m+1}\rangle]
\end{split}
\label{eqmvt2}
\end{equation}
and for ${\bold x} \neq {\bold y}$
\begin{equation}
\begin{split}
\frac{\partial}{\partial t}\langle a(t,{\bold x})a(t,{\bold y})\rangle= &\sum_{{\bold r}\neq {\bold 0}} D({\bold r})[\langle a(t,{\bold x})a(t,{\bold y}+{\bold r})\rangle+\langle a(t,{\bold x}+{\bold r})a(t,{\bold y})\rangle-\langle a(,t{\bold x})a(,t{\bold y})\rangle]\\
&-\lambda \ell [\langle a(t,{\bold x})a(t,{\bold y})^{m}\rangle-\langle a(t,{\bold x})^{m}a(t,{\bold y})\rangle]\\
\\
&+\mu k [\langle a(t,{\bold x})a(t,{\bold y})^{m}\rangle-\langle a(t,{\bold x})^{m}a(t,{\bold y})\rangle]
\end{split}
\label{eqmvt3}
\end{equation}
In particular, these equations close along the critical line (\ref{criticalline}).
The equation of motion of the two-time correlator reads 
\begin{equation}
\begin{split}
\frac{\partial}{\partial t}\langle a(t,{\bold x})a({s,\bold y})\rangle= &\sum_{{\bold n}\neq {\bold 0}} D({\bold n})[\langle a(t,{\bold x}+{\bold n})a(s,{\bold y})\rangle-\langle a(t,{\bold x})a(s,{\bold y})\rangle]\\
&-\lambda \ell \langle a(t,{\bold x})^{m}a(s,{\bold y})\rangle +\mu k\langle a(t,{\bold x})^{m}a(s,{\bold y})\rangle
\end{split}
\label{eqmvt4}
\end{equation}
together with the initial condition $\lim_{t\to s} \langle a(t,\vec{x}) a(s,\vec{y}) \rangle = \langle a(t,\vec{x})a(t,\vec{y}) \rangle$.

The relationship of these correlators with the average density and its variance are
given for diffusive transport by $\langle n(t,\vec{x})\rangle = \langle a(t,\vec{x})\rangle$ and $\langle n(t,\vec{x})^{2}\rangle = \langle a(t,\vec{x})^{2}\rangle + \langle a(t,\vec{x})\rangle$ \cite{Paes04a} and more generally by \cite{thesebau}, using also the definitions (\ref{gl:CRdef})
\BEQ \label{gl:nnCR}
\langle n(t,\vec{x}) n(s,\vec{x}+\vec{r})\rangle -\rho_0^2 = C(t,s;\vec{r}) + \rho_0 R(t-s,\vec{r})
\EEQ
In appendix~A, we generalise the proof of these relations to the case at hand. We shall show there that the second term in (\ref{gl:nnCR}) is merely generating corrections to the leading
scaling behaviour so that for our purposes, $C(t,s;\vec{r})$ can be interpreted as a connected density-density correlator. 

We begin the analysis of the equations of motion with the particle-density. For the BCPL, as well as for the BPCPL with $\sigma=0$, one can Fourier-transform eq.~(\ref{eqmvt1}), with the 
result 
\begin{equation}
\frac{\partial}{\partial t}\langle \wit{a}(t,{\bold q})\rangle
= -\frac{1}{2} 
\underbrace{\sum_{{\bold n}\neq {\bold 0}} 
\frac{D({\bold n})}{D}(1-e^{i{\bold q}\cdot{\bold n}})}_{=:\, \omega(\vec{q})} \;\;
\langle \wit{a}(t,{\bold q})\rangle+\frac{\sigma}{2} \langle \wit{a}(t,{\bold q})\rangle +\wit{h}(t,{\bold q})
\end{equation}
which defines the dispersion relation $\omega(\vec{q})$. If we can perform a continuum limit 
\begin{equation}
\omega({\bold q}) = 1-e^{-c\| {\bold q}\|^{\eta}} \underset{{\bold q} \rightarrow {\bold 0}}{\approx} c\| {\bold q}\|^{\eta}
\label{approxdisp}
\end{equation}
from which we read off the expected dynamical exponent $z=\eta$.  

Similarly, we calculate the response function by applying its definition (\ref{gl:CRdef}) to the equation of motion (\ref{eqmvt1}). This becomes in Fourier space (again, one must set $\sigma=0$ for the BPCPL) 
\begin{equation}
\frac{\partial}{\partial t} \wit{R}(t,s;{\bold q})= -\frac{1}{2}\omega({\bold q}) \wit{R}(t,s;{\bold q}) +\frac{\sigma}{2}  \wit{R}({t,s;\bold q}) +\delta(t-s)
\label{Rreso}
\end{equation}
Consequently, the real-space response function reads 
\begin{equation}
R({t,s;\bold r})=  
\frac{1}{(2\pi)^{d}} \underbrace{\int_{B} \D^{d}{\bold q} \: e^{i{\bold q}\cdot{\bold r}}
\: e^{-\frac{1}{2}\omega({\bold q})(t-s)}}_{=:b({\frac{t-s}{2},\bold r})} \; \exp\left({\frac{\sigma(t-s)}{2}}\right)\Theta(t-s)
\label{eqR}
\end{equation}
where $B$ is the Brillouin zone and the $\Theta$-function expresses causality. 
In particular, for long times the autoresponse becomes for $t>s$  
\BEQ
R({t,s;\bold 0}) \simeq  
e^{\frac{\sigma}{2}(t-s)} \left(\frac{t-s}{2}\right)^{-d/\eta}
\frac{1}{(2\pi)^{d}}\int_{0}^{\infty} \D^{d}{\bold q} \; e^{- \| {\bold q}\|^{\eta}c}
= \frac{2^{1-d}}{\pi^{d/2}}\frac{\Gamma\left(d/\eta\right)}{\eta \Gamma\left(d/2\right)} e^{\frac{\sigma}{2}(t-s)} \left(c\frac{t-s}{2}\right)^{-d/\eta}
\EEQ
For the correlation function, we shall assume that spatial translation-invariance holds and use the non-connected correlator
$F(t,s;\bold {r}):= \langle a(t,{\bold x})a(s,{\bold x}+{\bold r})\rangle$
and also introduce the control parameter
\begin{equation} \label{gl:alpha}
\alpha=\frac{\mu k(k+\ell)}{2D}
\end{equation}
As initial conditions, we shall use throughout the Poissonian distribution 
$F(0,0;{\bold r})=\rho_0^2$. Generalising slightly the calculations performed earlier 
for the BCPD and BPCPD \cite{Paes04a,Baum05}, we find for the BCPL the connected correlation function (on the critical line $\sigma=0$) 
\begin{equation} \label{Gcor}
C(t,s;{\bold r}) = \alpha \rho_{0}\int_{0}^{s}\D\tau \: b(\frac{t+s}{2}-\tau,{\bold r})
\simeq \alpha\rho_{0}\int_{0}^{s} \D \tau \:\int_{B} \frac{\D^{d}{\bold q}}{(2\pi)^{d}}
\exp\left({-c\| {\bold q}\|^{\eta}\left(\frac{t+s}{2}-\tau\right)}-\II \vec{r}\cdot\vec{q}\right)
\end{equation}
where the long-time limit $t,s\gg t_{\rm micro}$ and $t-s\gg t_{\rm micro}$ was also taken. 
See appendix~B for the computational details. 
For the bosonic pair-contact process BPCPL, we first consider the non-connected single-time correlation function $F(t,\vec{r})$ which satisfies the equation 
\begin{equation}
\begin{split}
F(t,{\bold r}) &=
\rho_{0}^{2} + \alpha\int_{0}^{t}\D\tau \: F(\tau,{\bold 0}) \int_{B} \frac{\D^{d}{\bold q}}{(2\pi)^{d}}e^{-\omega({\bold q})\left(t-\tau\right)}e^{\II{\bold q}\cdot{\bold r}}\\
&=\rho_{0}^{2} + \alpha\int_{0}^{t}\D\tau \: F(\tau,{\bold 0})\: b(t-\tau,{\bold r})
\end{split}
\label{FBPCP}
\end{equation}
For ${\bold r}=0$, this is a Volterra integral equation for $F({t,\bold 0})$ which may be solved by Laplace transformation as in (\cite{Paes04a}, \cite{Baum05}). The $F(t,{\bold r})$ is then directly obtained from (\ref{FBPCP}). Having found this, the two-time correlator is given by 
\BEQ \label{gl:Cts}
F(t,s,\vec{r})=\rho_0^2 + \alpha \int_0^s\D\tau F(\tau,\vec{0})b\left(\frac{t+s}{2}-\tau,\vec{r}\right)
\EEQ
The results of eqs. (\ref{Gcor},\ref{FBPCP},\ref{gl:Cts}) will form the basis of the subsequent analysis.
The essential difference with respect to the BCPD and BPCPD is the different form of the
dispersion $\omega(\vec{q})$. 

\section{Exact solution}

\subsection{Phase diagramme}
Since the total number of particles is conserved on average, information on the critical
behaviour comes by analysing the variance $\langle n(t,\vec{r})^2\rangle$. The resulting phase diagramme is presented in figure \ref{phasediagram}. There are two distinct phases separated by the critical line eq.~(\ref{criticalline}), namely (i) an absorbing phase for 
$\lambda > \mu$ with a vanishing particle-density in the steady-state, and (ii) an active phase for $\lambda < \mu$, where the particle-density diverges for large times. Along the critical line, the total mean particle-density is constant. The behaviour of the model along the
critical line, described by varying the control parameter $\alpha$ eq.~(\ref{gl:alpha}) is as follows: 

\begin{figure}[ht]
\begin{center}
\includegraphics[height=5cm]{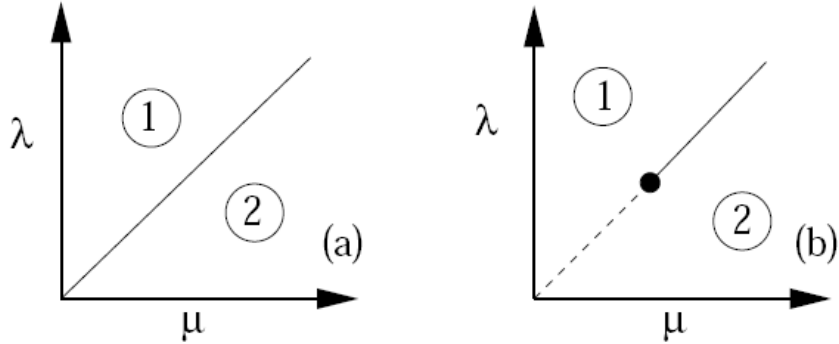}\\
\caption{Schematic phase-diagrammes for $d\neq0$ of (a) the BCPL and the BPCPL in $d\leq\eta$ and (b) the BPCPL in $d>\eta$. The absorbing region $1$ for $\lambda > \mu$, where the particle-density vanishes at large time, is separated by the critical line eq.~(\ref{criticalline}) from an active region for $\lambda < \mu$, where the particle-density diverges at large times. Along the critical line, the average time-dependent particle-density remains constant.\label{phasediagram}}
\end{center}
\end{figure}

\begin{itemize}
\item For the bosonic contact process BCPL, one has the same critical behaviour for all values of $\alpha$. On the other hand, the value of the space dimension $d$ is very important. 
If $d\leq \eta$, one has {\em clustering}: for long times, the particles are on average redistributed such as only a few, spontaneously selected, 
lattice sites contain particles while the others become essentially empty. 
For $d>\eta$, the long-time particle-density is spatially homogeneous. 

We can interpreted this result by using a generalised Polya theorem \cite{Paul99}, which states that for $d<\eta$ the L\'evy random walk is recurrent and hence cannot homogenise the system. On the other hand, when $d>\eta$, there is a non-zero probability that a L\'evy random walk does not return to its starting point. This is enough to make the system spatially homogeneous.
\item For the bosonic pair-contact process BPCPL, when $d<\eta$, there is always clustering. 
On the other hand, for $d>\eta$, there is a multi-critical point at $\alpha = \alpha_C$ such that the system is homogeneous for $\alpha<\alpha_C$ and that one has clustering for 
$\alpha >\alpha_C$. The critical point of this clustering transition is given by
\BEQ
\alpha_C =\left(2\int_{B} \frac{\D^d\vec{q}}{(2\pi)^d} (2\omega(\vec{q}))^{-1}\right)^{-1}
\EEQ
{}From this expression, one sees that $\alpha_C\to\infty$ if $d\to \eta$ \cite{Godr02,Baum06}. 
\end{itemize}

\subsection{Ageing exponents and scaling functions}
Using the explicit expressions from section~2, we can perform the long-time limit and
by comparing with the scaling forms eqs.~(\ref{CR1},\ref{CR2}) we can proceed to extract the
non-equilibrium critical exponents, which are listed in table~\ref{ageingexponent} and then the
scaling functions which, up to a normalisation factor, are given in table~\ref{scalingfunctions}. For the details of the calculation, see appendix~B. 

\begin{table}[tb]
\begin{center}
\begin{tabular}{|c|c||c|ccc|}
\hline
\multirow{2}{*}{} & Bosonic & \multicolumn{4}{c|}{Bosonic pair-contact process}\\
\cline{3-6}
 & contact process & $\alpha < \alpha_{C}$ & \multicolumn{3}{c|}{$\alpha = \alpha_{C}$}\\
\hline\hline
$a$ & $\frac{d}{\eta}-1$ & $\frac{d}{\eta}-1$ & \multicolumn{3}{c|}{$\frac{d}{\eta}-1$}\\
\hline
\multirow{2}{*}{$b$} &\multirow{2}{*}{$\frac{d}{\eta}-1$} & \multirow{2}{*}{$\frac{d}{\eta}-1$} & $0$ & if & $\eta < d < 2 \eta$ \\
 & & &  $\frac{d}{\eta}-2$ & if & $ d > 2 \eta$\\
\hline
$\lambda_{R}$ & $d$ & $d$ & \multicolumn{3}{c|}{$d$}\\
\hline
$\lambda_{C}$ & $d$ & $d$ & \multicolumn{3}{c|}{$d$}\\
\hline
$z$ & $\eta$ & $\eta$ & \multicolumn{3}{c|}{$\eta$}\\
\hline
\end{tabular}
\caption{Ageing exponents of the critical BCPL and BPCPL in different regimes. The results for the BCPL hold true for any dimension $d$, while for the BPCPL they only apply if $d>\eta$.\label{ageingexponent}}
\end{center}
\end{table}

\begin{table}[tb]
\begin{center}
\begin{tabular}{|c|c|c||c|c|}
\hline
\multicolumn{3}{|c|}{} & $f_{R}(y)$ & $f_{C}(y)$\\
\hline
\hline
\multicolumn{3}{|c|}{Bosonic contact process} & $(y-1)^{-d/\eta}$ & $(y-1)^{-d/\eta+1}-(y+1)^{-d/\eta+1}$ \\
\hline
\hline
Bosonic \multirow{4}{*}{} & $\alpha<\alpha_{C}$ & $d> \eta$ & $(y-1)^{-d/\eta}$ & $(y-1)^{-d/\eta+1}-(y+1)^{-d/\eta+1}$ \\
\cline{2-5}
pair contact & \multirow{3}{*}{$\alpha=\alpha_{C}$} & $\eta < d < 2 \eta$ & $(y-1)^{-d/\eta}$ & $(y+1)^{-d/\eta} \; _{2}F_{1}\left(\frac{d}{\eta},\frac{d}{\eta};\frac{d}{\eta}+1;\frac{2}{y+1}\right)$\\
\cline{3-5}
process & & \multirow{2}{*}{$d> 2\eta$} & \multirow{2}{*}{$(y-1)^{-d/\eta}$} & $ 2(d-2\eta)(y-1)^{1-d/\eta}$\\
 & & & & $-\eta(y-1)^{2-d/\eta}+\eta(y+1)^{2-d/\eta}$ \\ 
\hline
\end{tabular}
\end{center}
\caption{Scaling functions of the autoresponse and autocorrelation function of the critical bosonic contact and pair-contact processes. They are given up to a multiplicative factor.\label{scalingfunctions}}
\end{table}

Some comments are in order. 
\begin{itemize}
\item If $d>\eta$ and if furthermore $\alpha <\alpha_C$ in the BPCPL, the anticipated dynamical scaling (\ref{CR1},\ref{CR2}) holds true for both the BCPL and the BPCPL. Furthermore, since the exponents and the form of the scaling functions agree, the two models are in the same universality class.\footnote{In many respects, the behaviour of the BCPL found here is analogous to the ageing behaviour of a simple ferromagnet quenched onto its
critical point.}  
\item For $d<\eta$, we still find a dynamical scaling behaviour in the BCPL. We observe that the exponents $a=b$ now become {\em negative}. This reflects the eventual condensation of the particles which means that the fluctuations diverge for large times. 

For the BPCPL however, for $d<\eta$ or more generally $\alpha>\alpha_C$, no dynamical scaling is found. Rather, the variance diverges {\em exponentially} with time. The notion of ageing
as defined in the introduction no longer applies here. 
\item Finally, for the BPCPL with $d>\eta$ and at the tricritical point $\alpha =\alpha_C$,
the dynamical scaling (\ref{CR1},\ref{CR2}) holds true. The values of the 
non-equilibrium exponents are distinct from those of the universality class discussed above
and in particular, we note that $a \neq b$ although still $\lambda_C =\lambda_R$. 
In this respect, the absence of detailed balance has led in this case to an ageing behaviour
intrinsically different from the non-equilibrium dynamics of a ferromagnet (where $a=b$ as
well as $\lambda_C=\lambda_R$ always hold true). 
\item The response function $R=R(t-s;\vec{r})$ has a particularly simple form which actually obeys time-translation-invariance. However, since time-translation-invariance is not satisfied for the two-time correlators, the notion of ageing is still applicable.
\end{itemize}
If we identify the upper critical dimension as $d^*=\eta$, we see that our results can be mapped onto those \cite{Baum05} of the BCPD and BPCPD if one replaces $d/\eta = d/d^* \mapsto d/2$.

\section{Local scale-invariance}
We now inquire whether the forms of the scaling functions derived in the previous section can
be understood from some larger dynamical symmetry than mere dynamical scaling. 

Indeed, for simple ferromagnets undergoing phase-ordering kinetics after a quench from a
totally disordered initial state to the ordered phase with a temperature $T<T_c$, an extension
of dynamical scaling \cite{Bray94} to a {\em local} group of time-dependent scale-transformation related to
a subgroup of the Schr\"odinger group (without time translations) has been found \cite{Henk02,Henk07b}. In particular, it has been understood how to use the necessarily
projective representations of non-semisimple groups as the Schr\"odinger group in order 
to analyse the dynamical symmetries of the stochastic Langevin equations underlying these phenomena \cite{Pico04}. The essential tool in this kind of analysis are the celebrated 
Bargmann superselection rules \cite{Barg54}. Then both response as well as correlation
functions can be explicitly calculated and the results have been successfully tested in numerous models, see \cite{Henk07b} for a review. Since the Schr\"odinger group and its subgroups
only apply to physical situations where the dynamical exponent $z=2$, a generalisation to
arbitrary values of $z$ must be sought. 

\subsection{Background}
In this section, we shall use the BCPL with $z=\eta<2$ as an analytically treatable test case 
for such a possible extension. In order to do so, we shall first reformulate the problem
as a non-equilibrium field-theory using the Janssen-de Dominicis theory \cite{deDo76,Jans92,Howa97}. Starting from the master equation with the reaction rates eqs.~(\ref{gl:reac},\ref{gl:diff}), the creation and annihilation operators become related
in the continuum limit to the order-parameter field and a conjugate response field 
\BEA
\phi (t,{\bold r}) &:=a(t,{\bold r}) -\rho_0\\
\wit{\phi}(t,{\bold r}) &:= a^{\dagger}(t,{\bold r})-1
\EEA
such that $\langle \phi (t,{\bold r}) \rangle =\langle \wit{\phi} (t,{\bold r}) \rangle =0$. 
The action associated to the critical BCPL reads 
\BEQ
{\cal J}[\phi, \wit{\phi}] = \int \!\D{\bold R} \int \!\D u \; 
\left[\wit{\phi}(2{\cal M}\partial_u-\Delta^{\eta/2})\phi -\mu \wit{\phi}^{2} (\phi + \rho_0) \right] ={\cal J}_0[\phi,\wit{\phi}]+{\cal J}_b[\phi,\wit{\phi}]
\EEQ
where we have suppressed the arguments of $\phi(u,{\bold R})$ and $\wit{\phi}(u,{\bold R})$ under the integrals. The dimensionful `mass' $\cal M$ is related to a 
generalised diffusion constant. 
The action is decomposed into two parts: a so-called `deterministic' (noiseless) part 
\BEA
{\cal J}_0[\phi, \wit{\phi}] = \int \!\D{\bold R} \int \!\D u \; 
\left[\wit{\phi}(2{\cal M}\partial_u-\Delta^{\eta/2})\phi \right]
\EEA
and a `noise' part
\BEA
{\cal J}_b[\phi, \wit{\phi}] = -\int \!\D{\bold R} \int \!\D u \; \left[\mu \wit{\phi}^{2} (\phi + \rho_0) \right]
\EEA
Here the non-integral power of the Laplacian has to be interpreted as a fractional derivative
\cite{thesebau,Baum09}. In this formalism, correlators and response functions are found as
follows: 
\BEA
C(t_1,\ldots,t_n;\vec{r}_1,\ldots \vec{r}_n) &=& \langle \phi_1(t_1,{\bold r}_1)\ldots\phi_n(t_n,{\bold r}_n)\rangle 
\nonumber \\
R(t,s;\vec{r},\vec{r}') &=& \langle \phi(t,{\bold r})\wit{\phi}(s,{\bold r}')\rangle
\label{gl:CRdDJ}
\EEA
where the average of any observable $A$ is given by the functional integral 
$\langle A\rangle = \int \!\mathcal{D} \phi \mathcal{D} \wit{\phi}\: A[\phi,\wit{\phi}] 
e^{-{\cal J}[\phi, \tilde{\phi}]}$. As we have shown earlier for the case $z=2$ \cite{Pico04}, 
one considers first the `deterministic' part of the action, obtained from the
variation of ${\cal J}_0$ with respect to $\wit{\phi}$. 
The associated equation of motion
\BEQ \label{gl:LangBCPL}
{\cal S}\phi = \left( 2{\cal M}\partial_t - \Delta^{\eta/2} \right)\phi = 0
\EEQ
has a dynamical symmetry under the infinitesimal generators of local scale-transformations \cite{thesebau,Baum09}
\BEA
X_{-1}:=-\partial_t & \hspace{1.0 truecm} & \mbox{time-translation} \nonumber\\
X_0 := -t\partial_t -\frac{1}{\eta}(\vec{r}\cdot\partial_{\vec{r}})-\frac{\vec{x}}{\eta} & \hspace{1.0 truecm} & \mbox{dilatation} \nonumber\\
X_1:=-t^2\partial_t-\frac{2(x+\xi)}{\eta}t-\frac{2{\cal M}}{\eta^2} r^2\nabla^{2-\eta}_{\vec{r}}-\frac{2}{\eta}t(\vec{r}\cdot\partial_{\vec{r}}) & \hspace{1.0 truecm} & \mbox{generalised Schr\"odinger}\nonumber\\
-2\gamma (2-\eta)(\vec{r}\cdot\partial_{\vec{r}})\nabla^{-\eta}_{\vec{r}}-\gamma(2-\eta)(d-\eta)\nabla^{-\eta}_{\vec{r}} & \hspace{1 truecm} & \mbox{\hspace{0.5 truecm} transformation} \nonumber\\
Y^{(i)}_{-1/\eta}:= -\partial_{r_i} & \hspace{1.0 truecm} & \mbox{space rotation}\nonumber\\
Y^{(i)}_{-1/\eta+1}:= -t\partial_{r_i}-\frac{2{\cal M}}{\eta}  r_i\nabla^{2-\eta}_{\vec{r}}-\gamma \eta (2-\eta)\partial_{r_i} \nabla^{-\eta}_{\vec{r}} & \hspace{1.0 truecm} & \mbox{generalised Galilei} 
\label{gl:ielgen}\\
& \hspace{1.0 truecm} & \mbox{\hspace{0.5 truecm} transformation} \nonumber \\
R^{(i,j)} := - R^{(j,i)} \:=\: r_i\partial_{r_j} -r_j\partial_{r_i} & \hspace{1.0 truecm} & \mbox{rotation} \nonumber
\EEA
with $i,j=1,\ldots,d$, $\gamma$ is a dimensionful constant and $x,\xi$ are scaling dimensions. 
The fractional derivatives $\nabla_{\vec{r}}^{\alpha}$ with $\alpha\in\mathbb{R}$ 
are defined in \cite{thesebau,Baum09}. 
{}From the commutation relations \cite{thesebau,Baum09}
\BEA
{} \left[ X_n, X_{n'}\right] &=& (n-n') X_{n+n'} \nonumber \\
{} \left[ X_n, Y_m^{(i)} \right] &=& \left( \frac{n}{\eta}-m\right) Y_{n+m}^{(i)} \nonumber \\
{} \left[ Y_m^{(i)}, R^{(i,j)}\right] &=& Y_m^{(j)} \\
{} \left[ X_n, R^{(i,j)}\right] \:=\: 0 &;& \left[ Y_m^{(k)}, R^{(i,j)}\right]\:=\: 0 \;\; 
\mbox{\rm ~~if $k\ne i,j$}
\nonumber 
\EEA
it is clear that the explicitly specified generators span the complete algebraic structure. 
Furthermore $[{\cal S},{\cal X}]=0$ with all generators $\cal X$ of the above list (\ref{gl:ielgen}), with only two exceptions, namely
\BEA
{} \left[ {\cal S}, X_0\right] &=& -{\cal S} 
\nonumber \\
{} \left[ {\cal S}, X_1\right] &=& -2t{\cal S} -\frac{2{\cal M}}{\eta}\left( 
2(x+\xi) - (\eta-2+d) -\frac{\gamma}{2{\cal M}}\eta^2 (2-\eta)\right)
\EEA
Hence the infinitesimal generators (\ref{gl:ielgen}) map a solution of the `deterministic' 
equation ${\cal S}\phi=0$ onto another solution, provided $\phi$ the scaling dimensions 
$x,\xi$ of $\phi$ satisfy the condition
\BEQ
x+\xi = \frac{\eta-2+d}{2} + \frac{\eta^2 \gamma}{{4\cal M}}(2-\eta)
\EEQ
In order to be able to analyse the full theory, we first define the so-called 
`deterministic averages' $\langle A\rangle_0 := \int \!\mathcal{D} \phi \mathcal{D} \wit{\phi}\: A[\phi,\wit{\phi}] e^{-{\cal J}_0[\phi, \tilde{\phi}]}$ \cite{Pico04}. Consideration of
the iterated commutators of the generators $Y_m^{(i)}$ then shows that these deterministic
averages obey generalised Bargmann superselection rules \cite{thesebau,Baum09}
\begin{equation}\label{gl:Barg}
\left\langle \underbrace{\phi\,.\,.\,.\phi}_{n} \underbrace{\wit{\phi}\,.\,.\,.\wit{\phi}}_{m} \right\rangle_0 =0 , \hspace{1.0 truecm} \mbox{unless} \; n=m
\end{equation}
Now all building blocks are provided that we can adapt the treatment given earlier for
$z=2$ \cite{Pico04} to the present case.

\subsection{Application to the BCPL}
\begin{enumerate}
\item In order to compute a response function, one treats the `noise' part ${\cal J}_b$ of the
action as a perturbation. Using the relations (\ref{gl:CRdDJ}) and (\ref{gl:Barg}), we have 
\BEA
R(t,s;{\bold r},{\bold r^{\prime}}) &=& 
\left\langle \phi(t,{\bold r})\wit{\phi}(s,{\bold r^{\prime}}) e^{-{\cal J}_b[\phi,\wit{\phi}]}\right\rangle_0
\nonumber \\
&=& \left\langle \phi(t,{\bold r})\wit{\phi}(s,{\bold r^{\prime}}) \right\rangle_0
\nonumber \\
&=& R(t,s) \int_{\mathbb{R}^d} \frac{\D{\bold k}}{(2 \pi)^d} \vec{k}^{\beta_R}\mbox{exp}\left(\frac{\II ({\bold r}-{\bold r^{\prime}})\cdot {\bold k}}{(t-s)^{1/z}}-\alpha_R |{\bold k}| ^z \right)
\nonumber \\
&=:& R(t,s) {\cal F}^{(\alpha_R,\beta_R)}\left(\frac{{\bold r}-{\bold r^{\prime}}}{(t-s)^{1/z}}\right)
\EEA
and where we used the explicit form as determined from the covariance of the 'noiseless' response function.
This shows that the response function does not depend explicitly on the noise and may be found directly from the symmetries (\ref{gl:ielgen}) of the `deterministic' part alone. In particular, the form of the autoresponse function $R(t,s)=R(t,s;\vec{r},\vec{r})$ is determined by its covariance under the generator $X_1$ and reads
\BEA
R(t,s)= s^{-a-1} \left(\frac{t}{s}\right)^{1+a'-\lambda_R/z} \left(\frac{t}{s}-1\right)^{-1-a'}
\EEA
This result is in agreement with the scaling form (table \ref{scalingfunctions}) derived in section~3 with the
exponents $a'=a=d/\eta-1$, $\lambda_R=d$. 
\item We have for the space-time correlator, generalising the treatment of \cite{Baum06} to $z=\eta <2$, and following \cite{Baum06,Baum09}
\BEQ
C(t,s;\vec{r},\vec{r'}) = \left\langle \phi(t,\vec{r}) \phi(s,\vec{r'}) e^{-{\cal J}_b[\phi,\wit{\phi}]}\right\rangle_0
\EEQ
Expanding exponentials and using the Bargmann superselection rule, we find that the correlator is the sum of two terms $C_1(t,s;\vec{r},\vec{r'})$ and $C_2(t,s;\vec{r},\vec{r'})$
\BEQ
C_1(t,s;\vec{r},\vec{r'})=-2\mu\rho_0\int \D \vec{R} \int \D u \left\langle\phi(t,\vec{r})\phi(s,\vec{r'})\wit{\phi^2}(u,\vec{R})\right\rangle_0
\EEQ
and 
\BEQ
C_2(t,s;\vec{r},\vec{r'})=\mu^2 \int \D \vec{R} \D \vec{R'} \int \D u \D u' \langle\phi(t,\vec{r}) \phi(s,\vec{r'}) \Upsilon(u,\vec{R})\Upsilon(u',\vec{R'})\rangle_0
\EEQ
where $\Upsilon:=\wit{\phi^2}\phi$ is a composite field with a scaling dimension $x_{\Upsilon}= 2\wit{x} + x$. Furthermore, both terms of the correlator can be factorised as a product of response function \cite{thesebau} such that
\BEA \label{gl:C1}
\lefteqn{C_1(t,s;\vec{r},\vec{r'})=-2\mu\rho_0\int \D \vec{R} \int \D u \; \langle\phi(t,\vec{r})\wit{\phi}(u,\vec{R})\rangle_0\cdot\langle\phi(s,\vec{r'})\wit{\phi}(u,\vec{R})\rangle_0 } \nonumber \\
&=&-2\mu\rho_0 s^{-1-2a}\int_0^1 \D u\int_{\mathbb{R}^d}\D^d\vec{R} \; u^{-3-3a+\lambda_R/\eta} \left(\frac{t}{su}-1\right)^{-1-a} \left(\frac{1}{u}-1\right)^{-1-a} \nonumber\\
& &\times \left(\frac{t}{su}\right)^{1+a-\lambda_R/\eta}{\cal F}^{(\alpha_1,\beta_1)}\left(\frac{(\vec{r}-\vec{R})}{s^{1/\eta}(1-u)^{1/\eta}}\right){\cal F}^{(\alpha_2,\beta_2)}\left(\frac{(\vec{r'}-\vec{R})}{s^{1/\eta}(t/s-u)^{1/\eta}}\right)
\EEA
and
\BEA
\lefteqn{
C_2(t,s;\vec{r},\vec{r'})= 2\mu^2\int \D \vec{R} \int \D u \; \langle\phi(t,\vec{r})\Upsilon(u,\vec{R})\rangle_0\cdot\langle\phi(s,\vec{r'})\Upsilon(u,\vec{R})\rangle_0} \nonumber\\
&=& s^{1+\frac{4(x+\wit{x})}{\eta}}2 \mu^2\int_0^1 \D u\int_{\mathbb{R}^d}\D^d\vec{R} \; u^{\frac{4(x+\wit{x})}{\eta}} \left(\frac{t}{su}-1\right)^{-2((2\xi+x)+(2\xi_\Upsilon+x_\Upsilon))/\eta} \nonumber\\
& &\times \left(\frac{1}{u}-1\right)^{-2((2\xi+x)+(2\xi_\Upsilon+x_\Upsilon))/\eta} \left(\frac{t}{su}\right)^{2(2\xi_\Upsilon+x_\Upsilon-x)/\eta}\\
& &\times \left(\frac{t}{su}\right)^{1+a-\lambda_R/\eta}{\cal F}^{(\alpha_1,\beta_1)}\left(\frac{(\vec{r}-\vec{R})}{s^{1/\eta}(1-u)^{1/\eta}}\right){\cal F}^{(\alpha_2,\beta_2)}\left(\frac{(\vec{r'}-\vec{R})}{s^{1/\eta}(t/s-u)^{1/\eta}}\right)\nonumber
\EEA
Using $a=\frac{x+\wit{x}}{\eta}-1$, a dimensional analysis shows that 
\BEA
C_n(t,s;\vec{r},\vec{r'}) = s^{1-2nd/\eta} F_n(t/s,\vec{r}s^{-1/\eta},\vec{r}^{\prime}s^{-1/\eta})
\EEA
with $n=1,2$ and $F_n$ are scaling functions. It follows that in the scaling regime $s \rightarrow \infty$ (with the other scaling variables fixed) the term $C_2$ is negligible. It remains to see that $C_1$ is compatible with the expressions given in Table \ref{scalingfunctions}.
Using the identity \cite{thesebau},
\BEA
& &\int_{\mathbb{R}^d}\D\vec{R} \D\vec{R'} \; {\cal F}^{(\alpha_1,\beta_1)}\left(c\vec{R}+\vec{a}\right){\cal F}^{(\alpha_2,\beta_2)}\left(d\vec{R}+\vec{b}\right)g(\vec{R},\vec{R'}) \nonumber\\
&=&\int_{\mathbb{R}^d}\frac{\D^d\vec{k}}{(2\pi)^d}\int_{\mathbb{R}^d}\frac{\D^d\vec{q}}{(2\pi)^d} e^{\II\vec{a}\cdot\vec{k}+\II\vec{b}\cdot\vec{q}}\wit{g}(c\vec{k},d\vec{q})|\vec{k}|^{\beta_1}|\vec{q}|^{\beta_2}\exp(-\alpha_1|\vec{k}|^\eta)\exp(-\alpha_2|\vec{q}|^\eta)
\EEA
To recover the expression given in eq~(\ref{gl:C1}), we require $g(\vec{R},\vec{R'}) = \delta(\vec{R}-\vec{R'})$ and its Fourier transform reads $(2\pi)^d\delta(\vec{k}+\vec{q})$ . Then, we can easily find, with $\beta_1=\beta_2=\beta$ and $\alpha_1=\alpha_2$,
\BEA
C_1(t,s;\vec{r},\vec{r'})&=&
-2\mu\rho_0 s\int_0^1 \D u \; u^{-2-2a+2\lambda_R/\eta}(t/s-u)^{-1-a+\lambda_R/\eta} (1-u)^{-1-a+d/\eta} \nonumber\\
& &\times \int_{\mathbb{R}^d} \frac{\D^d\vec{k}}{(2\pi)^d} |k|^{2\beta}e^{\II(\vec{r}-\vec{r'})\cdot\vec{k}-2\alpha_1|\vec{k}|^\eta(t+s-2us)}
\EEA
Since $a=d/\eta-1$ and $\lambda_R=d$, the full space-time correlator reads
\BEA
C_1(t,s;\vec{r},\vec{r'})=-2\mu\rho_0 \int_0^s \D u \int_{\mathbb{R}^d} \frac{\D^d\vec{k}}{(2\pi)^d} |\vec{k}|^\beta e^{-2\alpha_1|\vec{k}|^\eta(t+s-2u)}e^{\II(\vec{r}-\vec{r^{\prime}})\cdot \vec{k}}
\EEA
where we identify $\beta=0$, $\alpha_1 = \frac{\II^\eta}{2\cal M}=\frac{c}{2}>0$ and ${\cal M} = \frac{\II^{\eta}}{c}$ to match to the expression~(\ref{Gcor}).
\end{enumerate}

Therefore, although the field-theory of BCPL is not free, the structure of its 'deterministic' part eq~(\ref{gl:LangBCPL}) is simple enough to explain the exact results as a manifestation of the local-scale-invariance.

\section{Passage time towards the ageing regime}
In the previous section, we have analysed the scaling behaviour of the BCPL and BPCPL.
Now, we analyse how from an initial state this scaling regime may be reached. Since the
BCPL and the BPCPL with $\alpha<\alpha_C$ are in the same universality class, we shall
limit ourselves in what follows to the BCPL or even to the BCPD. 
Our analysis generalises the previous study of
Zippold {\it et al.} \cite{Zipp00}, who examined this question for the low-temperature
spherical model and the spherical spin glass, in that we look at the critical case.

\begin{figure}[htb]
\begin{center}
\includegraphics[height=8cm]{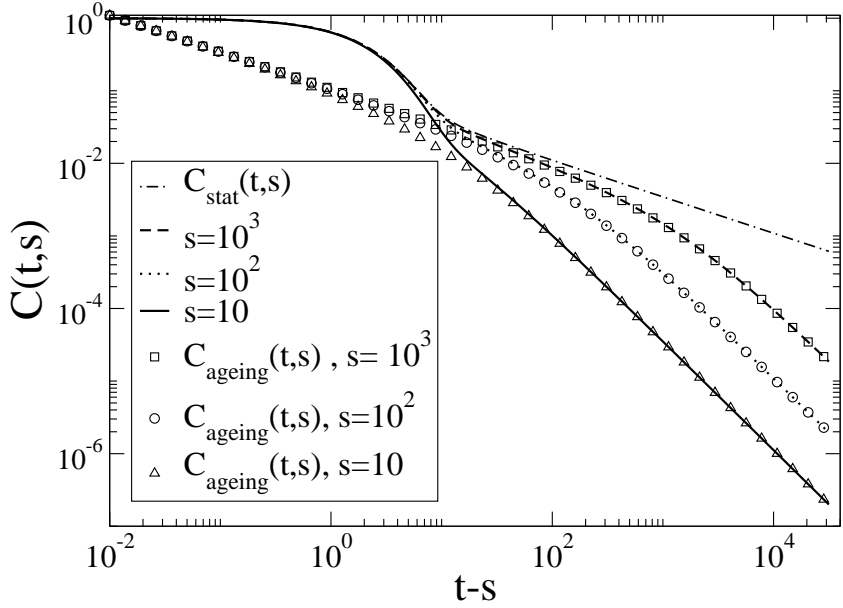}\\
\caption[Autocorrelator]{Autocorrelator $C(t,s)$ of the critical $3D$ BCPD ($\eta=2$) as a function of $t-s$, for several values of $s$. Full lines: complete autocorrelator. Dash-dotted line: stationary correlator eq.~(\ref{cstat}). Symbols: scaling contribution $C_{\rm ageing}(t,s)= s^{-b}f_C(t/s)$.}
\label{ccc}
\end{center}
\end{figure}

\begin{figure}[htb]\label{deltac}
\begin{center}
\includegraphics[height=8cm]{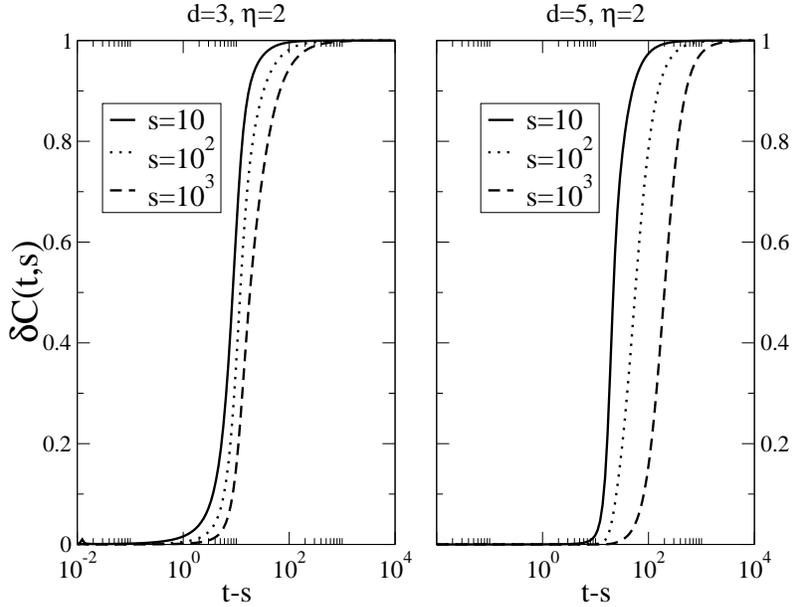}\\
\caption{Relative error $\delta C(t,s)$ as a function of time separation $t-s$ for different values of the waiting time $s$ in the BCPD. Left panel: $d=3$, right panel: $d=5$. \label{dbcp}}
\end{center}
\end{figure}

For a qualitative overview, we show in figure~\ref{ccc}, the autocorrelation function $C(t,s)$ as a function of time separation $t-s$. For small time-separations, the autocorrelator
remains close to the quasi-stationary, time-translationally-invariant form implied by the Poissonian initial conditions, before crossing over to the scaling regime with ageing behaviour when $t-s$ becomes large. In order to be able to define a precise measure of this cross-over,
we consider the following relative error
\begin{equation}
\delta C (t,s) = \left|\frac{C(t,s)-C_{stat}(t,s)}{C_{ageing}(t,s)-C_{stat}(t,s)}\right|
\label{relativeerror}
\end{equation}
where $C_{stat}(t,s)$ is the quasi-stationary correlation function
\begin{equation}
\label{cstat}
C_{stat}(t,s) = \alpha \rho_0 \int_{\frac{t-s}{2}}^\infty \!\D\tau \int _0 ^\pi \!\D q \, q^{d-1} \, \exp\left(\omega (q)\,\tau \right)
\end{equation}
and $C_{ageing}(t,s)=s^{-b} f_C(t/s)$ is the autocorrelator in the ageing regime, see eq.~(\ref{Gcor}). Since $0\leq \delta C(t,s)\leq 1$, we arbitrarily consider the system
to be in the stationary regime when $\delta C (t,s) <10\%$ and to be in the ageing regime when $\delta C (t,s) > 90\% $. In figure \ref{dbcp}, this relative error is plotted as a function of the separation time for different values of $s$. We observe that the passage from the quasi-stationary regime to the ageing regime occurs at relatively well-defined transition-times.

\begin{center}
\begin{figure}[htb]
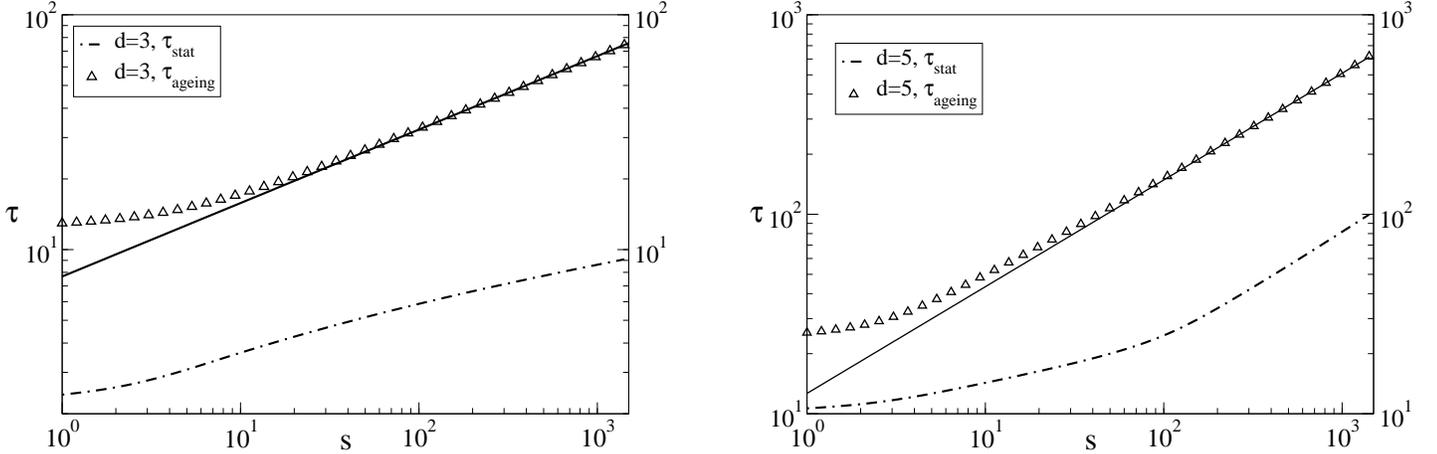

\begin{minipage}[c]{.45\linewidth}
\includegraphics[height=6cm]{taup.eps}\\
\end{minipage} \hfill
\begin{minipage}[c]{.45\linewidth}
\includegraphics[height=6cm]{taup2.eps}\\
\end{minipage}
\caption{Transition times $\tau_{\rm stat}(s)$ (lower curves) and $\tau_{\rm ageing}(s)$ (upper curves) as a function of $s$ of the BCPD, in $3D$ (left panel) and in $5D$ (right panel). The full lines give the linear fits: $\tau_{\rm ageing}(s)=7.7 \cdot s^{0.31}$ for $d=3$ and
$\tau_{\rm ageing}(s)=12.6 \cdot s^{0.53}$ for $d=5$.}
\label{resulttau}
\end{figure}
\end{center}

\begin{figure}[htb]
\begin{center}
\includegraphics[height=6.5cm]{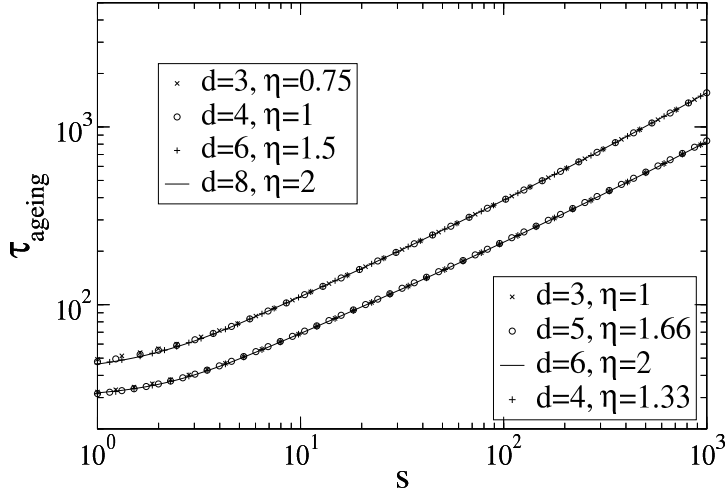}\\
\caption{Dependence of the transition time $\tau_{\rm ageing}(s)$ on the ratio $d/\eta$. 
The upper set of curves corresponds to $d/\eta=4$ and the lower one corresponds to $d/\eta=3$.}
\label{dsureta}
\end{center}
\end{figure}

Two transition times will be defined:
\begin{enumerate}
\item the system leaves the quasi-stationary regime at the time scale $\tau_{\rm stat}(s)$ which
is defined by
\BD
\delta C(\tau_{\rm stat}(s)+s,s) = 0.1
\ED
\item the system enters the ageing regime at the time scale $\tau_{\rm ageing}(s)$ which
is defined by
\BD
\delta C(\tau_{\rm ageing}(s)+s,s) = 0.9
\ED
\end{enumerate}
These transition times are shown as a function of the waiting time $s$ for several dimensions in figure~\ref{resulttau}. We find for sufficiently large $s$ the asymptotic behaviour
\BEA
\tau_{ageing}(t,s) = A\, s^{\zeta}
\label{taubehaviour}
\EEA
where $A$ depends on dimension and the {\em passage exponent} $\zeta<1$.
In principle, $\zeta$ should depend on both the dimension $d$ and the L\'evy parameter $\eta$. However, as illustrated in figure~\ref{dsureta}, $\zeta$ apparently depends merely on the
ratio $d/\eta$. Indeed, we obtain a neat collapse of the entire curves 
$\tau_{\rm ageing}(s)$  for different values of $d$ and $\eta$, but with the same value of $d/\eta$. Therefore, it is enough to
look at models with a fixed ratio of $d/\eta$ and for that reason most of our calculations 
were performed for the BCPD. 

\begin{figure}[bht]\label{fig:tausureta}
\begin{center}
\includegraphics[height=5.5cm]{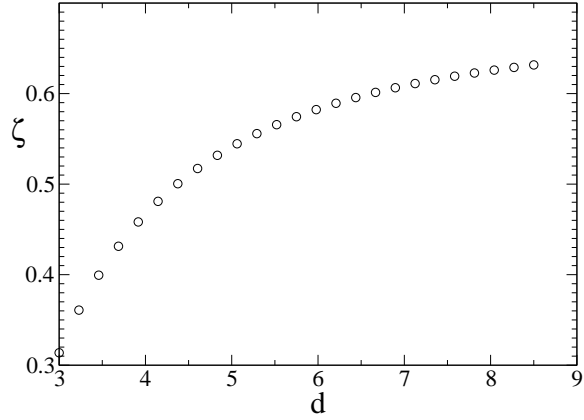}
\caption{Exponent $\zeta$ against dimension $d$ for the BCPD (where $\eta=2$ was used). If $d$ is replaced by $\eta d /2$, the results are also valid of the BCPL.}
\end{center}
\end{figure}

The dependence of $\zeta$ on the dimension $d$ is shown for $\eta=2$ in figure \ref{fig:tausureta} and in table~\ref{tableau1}. Analogously to the study of 
Zippold {\it et al.} \cite{Zipp00} for spherical ferromagnets and spherical spin glasses quenched to $T<T_c$, our results for
the {\em critical} passage exponent indicate that $\zeta<1$. In table~\ref{tableau1} we list some known values of $\zeta$ (the values for
the critical spherical ferromagnet are obtained in appendix~C). In conclusion, we find that the same qualitative behaviour for the
passage time between the quasi-stationary and the ageing regime, $\tau_{\rm ageing}(s)\sim s^{\zeta}$ with $0<\zeta<1$ apparently holds true both for critical as well as for non-critical systems. This form of the passage time is one of the ingredients 
for the derivation of admissible scaling forms of the two-time autocorrelator, which gives $C(t,s)=C_{\rm st}(t-s) +C_{\rm age}(t,s)$, where \cite{Andr06}
\BEQ
C_{\rm age}(t,s) = {\cal C}\left(\frac{h(t)}{h(s)}\right) \;\; , \;\;
h(t) = h_0 \exp\left[ \frac{1}{A} \frac{t^{1-\mu}-1}{1-\mu}\right]
\EEQ
where $\cal C$ is a scaling function, $\mu$ is a free parameter related to $\zeta$ and $h_0$ and $A$ are normalisation constants. However, while for quenches for $T< T_c$ the available evidence suggests that $\zeta$ should decrease with $d$, see table \ref{tableau1} \cite{Zipp00}, we find the opposite tendency for critical quenches, see figure \ref{fig:tausureta} and table \ref{tableau1}.

\begin{table}
\begin{center}
\begin{tabular}{|lc|ccl|l|} \hline
\multicolumn{2}{|l|}{model}         & condition      & ~~$d$~~ & ~$\zeta$~    & ~Ref.~ \\ \hline
\multicolumn{2}{|l|}{XY}            & $T=0$          & $1$     & $2/3$        & ~\cite{Godr03} \\ 
\multicolumn{2}{|l|}{spherical}     & $T<T_c$        & $>2$    & $4/(d+2)$    & ~\cite{Zipp00} \\ 
$p$-spin spherical glass   & $p=2$  & $T<T_c$        & --      & $4/5$        & ~\cite{Zipp00} \\
                           & $p=3$  & $T<T_c$        & --      & $\approx0.68$& ~\cite{Kim00} \\ \hline
\multicolumn{2}{|l|}{BCPD}          & $\lambda=\mu$  & $3$     & $0.31(1)$    & \\
                           &        &                & 3.5     & 0.40(1)    & \\
                           &        &                & 4       & 0.47(1)    & \\ 
                           &        &                & 5       & 0.53(1)    & \\ 
                           &        &                & 6       & 0.58(1)    & \\ 
\multicolumn{2}{|l|}{spherical}     & $T=T_c$        & $3$     & $0.33(1)$  & \\
                           &        &                & $5$     & $0.40(1)$  & \\ \hline
\end{tabular} 
\end{center}
\caption{Values of the passage exponent $\zeta$ defined in eq.~(\ref{taubehaviour}) 
for several models, either in the coexistence phase or at criticality.\label{tableau1}}
\end{table}

\begin{figure}[htb]
\begin{center}
\includegraphics[height=8cm]{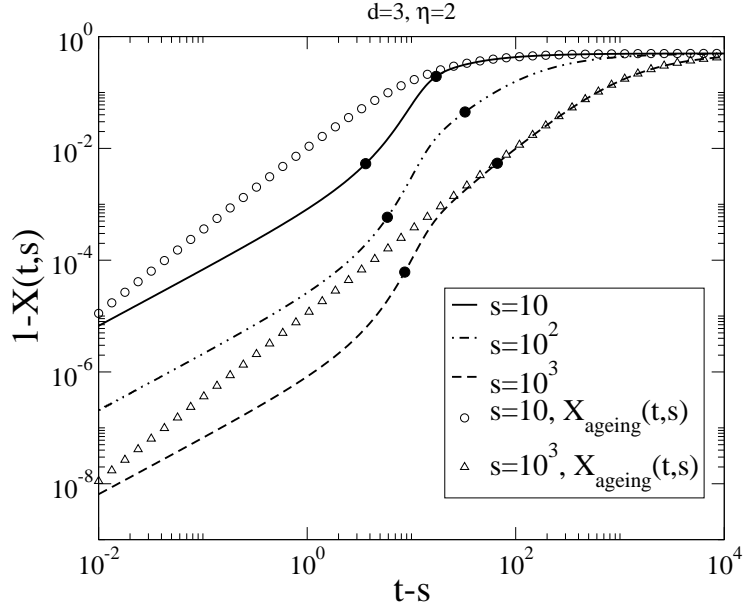}
\end{center}
\caption{Fluctuation-dissipation ratio $X(t,s)$ (full lines) and the ageing limit $X_{\rm ageing}(t,s)$ (symbols) as a function of the time separation $t-s$ for different value of $s$ in the $3D$ BCPD.}
\label{rfd}
\end{figure}

The passage towards the ageing regime can still be understood heuristically in a different way. At equilibrium, the fluctuation-dissipation theorem (FDT) describes the expected size of the fluctuations, given the response to an external perturbation. If the FDT is broken in ageing phenomena, the inverse fluctuation-dissipation ratio (FDR) describes by how much this expectation is larger than the actually found result, for a given value of the temperature $T$
\cite{Kurch}. Since for the BCPL, the ageing exponents $a=b$ are equal, we can define an analogue of a fluctuation-dissipation ratio
\begin{equation}
X(t,s) = \frac{\frac{\partial C}{\partial s}(s,s)}{R(s,s)} \, \frac{R(t,s)}{\frac{\partial C}{\partial s}(t,s)}
\end{equation}
but where now the initial FDR plays the role the temperature in system with an equilibrium stationary state. From our exact solution of the BCPL, we find, in the ageing regime
\begin{equation}
X_{\rm ageing}(t,s) = \frac{1}{1+ \left( \frac{t-s}{t+s}\right) ^{d/\eta}}
\end{equation}
In figure~\ref{rfd}, we show $X(t,s)$ and $X_{\rm ageing}(t,s)$ as a function of the time separation $t-s$ for different waiting times $s$. We can distinguish three different regimes:
\begin{enumerate}
\item a quasi-stationary regime with microscopic relaxation for $t-s \ll \tau_{\rm stat}(s)$
\item a non-analytic transition regime for $\tau_{\rm stat}(s) \ll t-s \ll \tau_{\rm ageing}(s)$
\item the ageing regime when $t-s \gg \tau_{\rm ageing}(s)$. 
\end{enumerate}
The bold circles in figure~\ref{rfd} indicate the transitions between these regimes. We see that
the FDR is still very close to unity when the quasi-stationary regime is left and not too far
from its limit value $X_{\infty}=\frac{1}{2}$ upon entering the ageing regime. 

\section{Conclusions}
In analysing the non-equilibrium dynamical scaling of two `bosonic' particle-reaction models with 
L\'evy-flight transport of individual particles, characterised by the parameter $\eta$, instead of diffusive transport, 
we have obtained the following results:
\begin{enumerate}
\item We have defined the `bosonic' contact and pair-contact processes BCPL and BPCPL. Because of the
absence of a `fermionic' constraint limiting the number of particles per site, the models are exactly
solvable, either in the full parameter space as the BCPL or along the critical line as the BPCPL. Since the
global particle number is conserved on the critical line, information on the nature of the phase transition
comes from analysing the variance of the particle distribution. This shows that the BCPL and BPCPL may give rise
to a condensation of the particles on essentially a single lattice site. 
\item Starting from an initial condition far from the stationary state, the two-time correlation and response functions
obey the same kind of dynamical scaling and ageing, see eq.~(\ref{CR1},\ref{CR2}) as already found in other systems without detailed 
balance. The explicit results for the non-equilibrium exponents and the scaling functions are given in tables~\ref{ageingexponent} and \ref{scalingfunctions} for arbitrary values $0<\eta<2$. 

For $\eta=2$ our results include those of the diffusive case (models BCPD and BPCPD \cite{Houc02,Paes04a,Baum05}) as special cases. 
\item We have used these explicit results to test a formulation of local scale-invariance for a dynamical exponent $z=\eta< 2$.
Extensions of global dynamical scaling to a more local form in non-equilibrium systems must be capable of treating the dynamical
symmetries of {\em stochastic} Langevin equations, which is done by attempting to perturb around the noiseless, `deterministic'
part of the model, which in the case of the BCPL studied here takes the simple Markovian form 
$\bigl( 2{\cal M}\partial_t - \Delta^{\eta/2}\bigr)\phi=0$, where $\cal M$ is a dimensionful constant. Local scale-invariance shows how generalised Bargmann superselection rules can be derived for arbitrary values of
the dynamical exponent such that the perturbation series naturally truncates. Our analysis shows how the procedure may be extended
to systems without detailed balance and in this sense generalises earlier studies performed on long-ranged ferromagnets \cite{Baum06b,Dutta08}. 
\item Finally, we used the exact solution to follow precisely the passage between the quasi-stationary regime and the late-time
ageing regime. In analogy with earlier studies on this passage in non-critical systems \cite{Zipp00,Godr03}, our results suggest that the 
passage time depends on the waiting time $s$ also for {\em critical} ferromagnets, as well as for systems without detailed balances, 
as $\tau_{\rm ageing}(s)\sim s^{\zeta}$, with $0<\zeta<1$. 
\end{enumerate}
All in all, the explicit study of the exactly solvable BCPL and BPCPL processes has permitted us to control and to confirm
several assumptions, usually admitted in studies of non-equilibrium dynamical scaling, and hence to extend their range of applicability.

\appsection{A}{Proof of eq. (\ref{gl:nnCR})}
We generalise Baumann's proof \cite{thesebau}, 
valid for diffusive motion of single particles, to the
case of long-range jumps. We work in the quantum hamiltonian/Liouvillian formalism. By the property $\langle s| a^{\dagger}(x)= \langle s|$, we can verify that along the critical line 
\begin{equation}
\partial_t \langle s| a(\vec{x})= \langle s| [a(\vec{x}),H] = \sum_{\bold n \neq 0} \langle s| D(\vec{n}) \left((a(\vec{x}+\vec{n}) -a(\vec{x})\right)
\end{equation}
The Fourier transform is defined as $\langle s| \wit{a}({\bold k}) = \sum_{\bold x} e^{-\II {\bold k.x}} \langle s| a({\bold x })$, therefore 
\BEA
\langle s| \wit{a}(t,{\bold k})= \exp \left(-\frac{1}{2} \omega({\bold k})(t-s)\right) \langle s|\wit{a}(s, {\bold k})
\EEA
where $\omega({\bold k})$ is the dispersion relation. Using Fourier's theorem  $\langle s| a({\bold x}) = (2\pi)^{-d} \int_{B} d{\bold k} \; e^{-i {\bold k.x}} \langle s| \tilde{a}({\bold kx }) $, we have for all times $s \leq t$ 
\BEA
\begin{split}
\langle s| a(t,{\bold x}) &= \sum_{\bold y} \int_B \frac{\D{\bold k}}{(2\pi)^d}\; e^{\II\vec{k}\cdot(\vec{x}-\vec{y})}\exp\left(-\frac{1}{2} \omega({\bold k})(t-s)\right) \langle s|a(s, \bold {y})\\
&= \sum_{\bold y} R(t-s;\vec{x}-\vec{y}) \langle s| a(s, {\bold y})
\end{split}
\label{genC}
\EEA
where we have used the expression (\ref{eqR}) of the response function.
Next, we apply (\ref{genC}) to the two-point correlation function 
\BEA \label{gl:A:FF}
F(t,s;{\bold r}) = \langle a(t,{\bold x})a(s,\bold{x}+\vec{r})\rangle = \sum_{\bold y} R(t-s; \bold{x}-\vec{y})F(s,\bold{r}+\vec{x}-\vec{y})
\EEA
We also calculate the density-density correlator and we find
\BEA
\begin{split}
\langle n(t,{\bold x})n(s,\bold{x}+\vec{r})\rangle &= \langle a^{\dagger}(t,{\bold x})a(t,{\bold x})a^{\dagger}(s,\bold {x}+\vec{r})a(s,\bold {x}+\vec{r})\rangle\\
&=\sum_{\bold y} R(t-s;\bold {x}-\vec{y}) \langle a(t,{\bold x})a^{\dagger}(s,\bold {x+r})a(s,\bold {x+r})\rangle\\
&=\sum_{\bold y} R(t-s;\bold {x}-\vec{y})F(s,\bold {r}+\vec{x}-\vec{y})+R(t-s;{\bold r})\rho_0\\
&=F(t,s;\vec{r}) +R(t-s;{\bold r})\rho_0
\end{split}
\EEA
where we have used the time-dependent commutator of $a$ and $a^{\dagger}$ and also applied
(\ref{gl:A:FF}). Going back to the definition (\ref{gl:CRdef}) the assertion
(\ref{gl:nnCR}) follows. 

Finally, using the scaling forms (\ref{CR1},\ref{CR2}) together with the exponents
of table~\ref{ageingexponent}, we explicitly see that the second term in eq.~(\ref{gl:nnCR})
becomes negligible compared to the first one in all cases.

\appsection{B}{Analysis of the correlators}
We discuss those aspects in the calculation of the correlators which go beyond the
local BCPD and BPCPD models \cite{Paes04a,Baum05}. First, we reconsider the Green's function
$b(t,\vec{r})$, see eq.~(\ref{approxdisp}). In the ageing regime ($t,s \gg 1$ et $t-s \gg 1 $), only small values of ${\bold q}$ will contribute. Therefore, 
\begin{equation}
b(t,{\bold r})=\frac{1}{(2\pi)^{d}} \int_{B} \!\D^{d}{\bold q}\: 
e^{\left(\II{\bold q}\cdot{\bold r}- \omega({\bold q}) t\right)}
\underset{t\rightarrow \infty}{=}\frac{t^{-d/\eta}}{(2\pi)^{d}} \int_{\mathbb{R}^d} \!\D^{d}{\bold q}\: e^{\left(\II{\bold q}\cdot{\bold r}t^{-1/\eta}-\| {\bold q}\|^{\eta}c \right)}
\label{brep}
\end{equation}
{}From this, we identify the scaling variable ${\bold \xi}={\bold r}t^{-1/\eta}$ and read off
the dynamical exponent $z=\eta$.

For the explicit calculation of the correlator in the BCPL, we merely have to evaluate 
eq.~(\ref{Gcor}). For the autocorrelator, we find
\begin{itemize}
\item Case $d=\eta$:
\begin{equation}
C(t,s;{\bold 0}) =  \alpha\rho_{0}B_0\int_{\frac{t-s}{2}}^{\frac{t+s}{2}}\D\tau \: \tau = \alpha\rho_{0}B_0 \textrm{ ln} \left(\frac{t/s+1}{t/s-1}\right) = f_{C}(t/s)
\label{logCBCP}
\end{equation}
where $B_0= \frac{2^{1-d}\Gamma(d/\eta)}{\eta c^{d/\eta} \pi^{d/2}\Gamma(d/2)}$
and we read off the exponents $b=0$ and $\lambda_{C}=d$.
\item Case $d\ne\eta$:
\BEA
C(t,s;\vec{0}) &=& 
 \frac{ \alpha\rho_{0}B_0}{1-d/\eta} \left(\left(\frac{t+s}{2}\right)^{1-d/\eta}-\left(\frac{t-s}{2}\right)^{1-d/\eta}\right)
\nonumber \\
&\simeq&  \frac{ \alpha\rho_{0}B_0}{2^{-d/\eta}}\: s^{1-d/\eta} \left(\frac{t}{s}\right)^{-d/\eta}
\EEA
and in the second line we looked at the asymptotic form if $t/s\to \infty$. We read off
the exponent $b=-1+d/\eta$ and $\lambda_C=d$. 
\end{itemize}
The space-time-dependent scaling functions can be found similarly. We merely quote the result:
\BEQ
C(t,s;\vec{r}) = \alpha \rho_0 B_0  \left(C_{-1}\left(\frac{\vec{r}}{\left(\frac{t-s}{2}\right)^{1/\eta}}\right)\left(\frac{t-s}{2}\right)^{1-d/\eta} - C_{-1}\left(\frac{\vec{r}}{\left(\frac{t+s}{2}\right)^{1/\eta}}\right)\left(\frac{t+s}{2}\right)^{1-d/\eta} \right)
\EEQ
where the function $C_{-1}$ reads
\BEQ
C_{-1}(\vec{r}t^{-1/\eta})= t^{d/\eta} B_0^{-1}\int_{\mathbb{R}^d} \D^d\vec{q} \;\; e^{\II \vec{q}\cdot\vec{r}} \; e^{-|\vec{q}|^\eta t}\left(c|q|^\eta\right)^{-1}
\EEQ
For the BPCPL, using Laplace transform on eq~(\ref{FBPCP}) and applying a similar analysis than in \cite{Godr02}, we evaluate $F(t,\vec{0})$ in the different cases and then use the eq~(\ref{gl:Cts}) to compute the connected correlator. The most interesting cases are $\alpha \leq \alpha_C$ and $d>\eta$
\begin{enumerate}
\item \underline{$\alpha < \alpha_C$ and $d>\eta$:} here, $F(t,\vec{0}) = \frac{\alpha_C \rho_0^2}{\alpha_C-\alpha}$ and the scaling function reads
\begin{equation}
f_{C}(y)=\frac{\alpha\alpha_C \rho_{0}^{2}B_0}{(\alpha-\alpha_C) \left(\frac{2d}{\eta}-2\right)2^{-d/\eta}}\left[(y+1)^{1-d/\eta}-(y-1)^{1-d/\eta}\right]
\label{G0BPCPech}
\end{equation}
\item \underline{$\alpha = \alpha_C$:}\\
For $\eta<d<2\eta$, $F(t,\vec{0})=\frac{\rho_0^2}{\alpha_C B_0 |\Gamma(1-d/\eta)|\Gamma(d/\eta)}t^{d/\eta-1}$ and the scaling function is given by
\begin{equation}
f_{C}(y)=\frac{\rho_{0}^{2}}{|\Gamma(1-d/\eta)|\Gamma(d/\eta)d/\eta}\left(\frac{y+1}{2}\right)^{-d/\eta} \;_{2}F_{1}\left(\frac{d}{\eta},\frac{d}{\eta};\frac{d}{\eta}+1;\frac{2}{y+1}\right)
\end{equation}
For $d>2\eta$
\begin{equation}
\begin{split}
f_{C}(y)=\frac{\rho_{0}^{2}B_0\eta}{A_2 2^{-d/\eta+2}(d-\eta)\left(d-2\eta\right)}\left[2(d-2\eta)\left(y-1\right)^{1-d/\eta }-\eta(y-1)^{2-d/\eta}+\eta(y+1)^{2-d/\eta}\right]
\end{split}
\label{G0BPCPech2}
\end{equation}
where $A_2=\left(2\int_{B} \frac{\D^d\vec{q}}{(2\pi)^d} (2\omega(\vec{q}))^{-2}\right)^{-1}$.
\item \underline{$\alpha > \alpha_{C}$ or $d < \eta$:} $F(t,\vec{0})$ shows an exponential behaviour and we do not have scaling behaviour in these cases.
\end{enumerate}

The space-time-dependent scaling functions can be found from eq~(\ref{gl:Cts}).
\begin{enumerate}
\item \underline{$\alpha < \alpha_C$ and $d>\eta$:} 
\begin{equation}
C(t,s;\vec{r}) = \alpha \rho_0 B_0  \left(C_{-1}\left(\frac{\vec{r}}{\left(\frac{t-s}{2}\right)^{1/\eta}}\right)\left(\frac{t-s}{2}\right)^{1-d/\eta} - C_{-1}\left(\frac{\vec{r}}{\left(\frac{t+s}{2}\right)^{1/\eta}}\right)\left(\frac{t+s}{2}\right)^{1-d/\eta} \right)
\end{equation}
\item \underline{$\alpha = \alpha_C$:}\\
For $\eta<d<2\eta$
\begin{equation}
C(t,s;\vec{r})= \frac{\rho_0^2}{|\Gamma(1-d/\eta)|\Gamma(d/\eta)} \left(\frac{\frac{t}{s}+1}{2}\right)^{-d/\eta} \sum_{n=0}^{\infty} \frac{C_n\left(\vec{r}\left(\frac{t+s}{2}\right)^{-1/\eta}\right)\left(\frac{\frac{t}{s}+1}{2}\right)^n}{n!(n+d/\eta)}
\end{equation}
where $C_n$ is defined as
\BEQ
C_{n}(\vec{r}t^{-1/\eta})= t^{d/\eta} B_0^{-1} \int_{\mathbb{R}^d} \D^d\vec{q} \;\; e^{\II \vec{q}\cdot\vec{r}} \; e^{-|\vec{q}|^\eta t} \left(c|q|^\eta\right)^n
\EEQ
For $d>2\eta$
\begin{equation}
C(t,s;\vec{r})=\frac{\rho_0^2}{4A_2} B_0s^{2-d/\eta}\left(\frac{\frac{t}{s}+1}{2}\right)^{-d/\eta} \sum_{n=0}^{\infty} \frac{C_n\left(\vec{r}\left(\frac{t+s}{2}\right)^{-1/\eta}\right)\left(\frac{\frac{t}{s}+1}{2}\right)^n}{n!(n+2)}
\label{G0BPCPech3}
\end{equation}
\end{enumerate}

\appsection{C}{Critical spherical model}

We analyse the passage towards the ageing regime in the 
critical spherical model. As introduced by Berlin and Kac \cite{Berl52}, 
its spins $S_{\vec{x}}\in\mathbb{R}$ are real variables on a hypercubic lattice $\Lambda$ in $d$ dimensions which obey the
so-called spherical constraint 
\BEQ
\sum_{\vec{x}\in\Lambda} S^2_{\vec{x}}=\cal{N}
\EEQ
where $\cal{N}$ is the total number of spins. The dynamics is given by the stochastic Langevin equation
\BEQ
\partial_tS(t,\vec{r}) = \nabla_{\vec{r}}^2 S(t,\vec{r}) + \frac{1}{2}\left(\frac{\D}{\D t} \ln g(t)\right) S(t,\vec{r}) +\eta(t,\vec{r})
\EEQ
where $\eta(t,\vec{r})$ is a centred gaussian noise such that $\langle \eta(t,\vec{r})\eta(t',\vec{r'})\rangle=2T_c\delta(t-t')\delta(\vec{r}-\vec{r'})$ and $g(t)$ acts as a 
Lagrange multiplier to enforce the spherical constraint, see below. We only quote from the well-known solution \cite{Godr00b} 
those results we need. The autocorrelation reads (we use an infinite-temperature initial state throughout) 
\BEQ
C(t,s) = \langle S(t,\vec{r}) S(s,\vec{r}) \rangle = \frac{1}{\sqrt{g(t)g(s)}}\left(f\left(\frac{t+s}{2}\right)+2T_c \int_0^s \D t' f\left(\frac{t+s}{2}-t'\right) g(t') \right)\\
\EEQ
where $g(t)$ is given as the solution of the Volterra integral equation
\BEQ \label{gl:C:volt}
g(t)=f(t) +2T_c\int_0^t \D t' f(t-t')g(t')
\EEQ
and $f(t)$ reads for short-range interactions ($I_0$ is a modified Bessel function)
\BEQ
f(t)=\int_ B \frac{\D\vec{q}}{(2\pi)^d}e^{-2\omega(\vec{q})t}= \left(e^{-4t}I_0(4t)\right)^d
\EEQ 
Finally, the critical temperature $T_c$ is given by
\BEQ
T_c(d)= \left(2\int_0^{\infty} \D t \left(e^{-4t}I_0(4t)\right)^d \right)^{-1}
\EEQ
We also recall the scaling functions of the autocorrelator in the ageing regime \cite{Godr00b,Godr02}
\BEA
f_C(y) &= & T_c \frac{4(4\pi)^{-d/2}}{(d-2)(y+1)} y^{1-d/4} (y-1)^{1-d/2} \hspace{3 truecm} 2<d<4, \nonumber \\
f_C(y) &= & T_c \frac{2(4\pi)^{-d/2}}{d-2} \left((y-1)^{1-d/2}-(y+1)^{1-d/2}\right) \hspace{1.8 truecm} d>4
\label{gl:C:asymp}
\EEA

\begin{figure}[htb]
\begin{center}
\includegraphics[height=6cm]{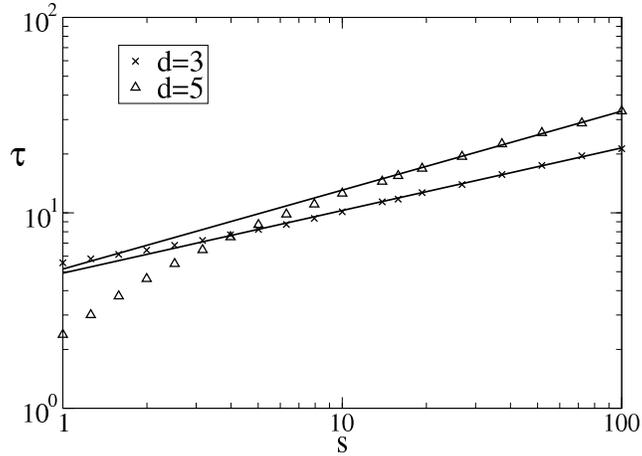}\\
\caption{Passage time towards the ageing regime in the spherical model at $T=T_c$,as a function of the waiting time $s$, for different values of the dimension $d$. The full lines give the linear fits: $\tau_{\rm ageing}(s)=4.9\cdot s^{0.32}$ for $d=3$ and
$\tau_{\rm ageing}(s)=5.2 \cdot s^{0.40}$ for $d=5$.}
\label{tauspherical}
\end{center}
\end{figure}

As in the main text, we define the passage times between the quasi-stationary and the ageing regimes. 
In figure~\ref{tauspherical}, we show
$\tau_{\rm ageing}(s)\sim s^{\zeta}$ as a function of $s$. The full curves indicate the numerical
solution of the Volterra integral equation (\ref{gl:C:volt}) whereas the symbols were 
obtained from the asymptotic forms (\ref{gl:C:asymp}). Hence $\tau_{\rm ageing}(s)$ shows
the same kind of qualitative behaviour as the BCPL and the resulting values of $\zeta$ are included in table~\ref{tableau1}. 

\noindent
{\bf Acknowledgements}
XD acknowledges the support by the Centre National de la Recherche Scientifique (CNRS) through grant no. 101160. We thank S.B. Dutta for useful discussions.

\newpage 

\end{document}